\documentclass[journal=nalefd,manuscript=article]{achemso}
\usepackage[version=3]{mhchem} % Formula subscripts using \ce{}

\usepackage{graphicx}% Include figure files
\usepackage{dcolumn}% Align table columns on decimal point
\usepackage{bm}% bold math
\usepackage{hyperref}% add hypertext capabilities
\usepackage{color,soul}
\title{Spin-orbit torques in NbSe$_2$/Permalloy bilayers}

\author{Marcos H. D. Guimar\~aes}
\affiliation{Laboratory of Atomic and Solid State Physics, Cornell University, Ithaca – NY, 14853, USA}
\alsoaffiliation{Kavli Institute for Nanoscale Science, Cornell University, Ithaca – NY, 14853, USA}
\author{Gregory M. Stiehl}
\affiliation{Laboratory of Atomic and Solid State Physics, Cornell University, Ithaca – NY, 14853, USA}
\author{David MacNeill}
\affiliation{Laboratory of Atomic and Solid State Physics, Cornell University, Ithaca – NY, 14853, USA}
\author{Neal D. Reynolds}
\affiliation{Laboratory of Atomic and Solid State Physics, Cornell University, Ithaca – NY, 14853, USA}
\author{Daniel C. Ralph}
 \email{dcr14@cornell.edu}
\affiliation{Laboratory of Atomic and Solid State Physics, Cornell University, Ithaca – NY, 14853, USA}
\alsoaffiliation{Kavli Institute for Nanoscale Science, Cornell University, Ithaca – NY, 14853, USA}

\begin{document}


\begin{abstract}
We present measurements of current-induced spin-orbit torques generated by NbSe$_2$, a fully-metallic transition-metal dichalcogenide material, made using the spin-torque ferromagnetic resonance (ST-FMR) technique with NbSe$_{2}$/Permalloy bilayers.
In addition to the out-of-plane Oersted torque expected from current flow in the metallic NbSe$_{2}$ layer, we also observe an in-plane antidamping torque with torque conductivity $\sigma_{S} \approx 10^{3} (\hbar / 2e)$($\Omega$m)$^{-1}$ and indications of a weak field-like contribution to the out-of-plane torque oriented opposite to the Oersted torque. 
Furthermore, in some samples we also measure an in-plane field-like torque with the form $\hat{m} \times \hat{z}$, where $\hat{m}$ is the Permalloy magnetization direction and $\hat{z}$ is perpendicular to the sample plane.
The size of this component varies strongly between samples and is not correlated with the NbSe$_{2}$ thickness.
A torque of this form is not allowed by the bulk symmetries of NbSe$_{2}$, but is consistent with symmetry breaking by a uniaxial strain that might result during device fabrication.
\end{abstract}

%\pacs{72.80.Vp, 72.25.-b, 85.75.Hh}% PACS, the Physics and Astronomy
%\keywords{Transition Metal Dichalcogenides, spin orbit torque, spin orbit coupling, van der Waals}  %Use showkeys class option if keyword
\maketitle

%%%%%%%%%%%%%%%%%%%%%%%%%%%%%%%%%%%%%%%%%%%%%%%%%%%%%%%%%%%%%%%%%%%%%%%%%%%%%%%%%%%%%%%
%\section{Introduction}

Current-induced spin torques generated by materials with large spin-orbit coupling (SOC), such as heavy metals \cite{Liu2011,Miron2011,Pai2012} and topological insulators \cite{Fan2014,Mellnik2014}, are candidates to enable a new generation of efficient non-volatile magnetic memories.
Several research groups have recently considered the possibility that some 2-dimensional (2D) materials might also be used as sources of  spin-orbit torque (SOT) \cite{Zhang2016,Shao2016,Cheng2015,MacNeill2016,MacNeill2017}.
For example, 2D transition metal dichalcogenides \cite{Manzeli2017} (TMDs) can possess strong SOC and are easily incorporated into device heterostructures with clean, atomically-precise interfaces. Initial studies of the SOT originating from the TMD semiconductors MoS$_{2}$ \cite{Shao2016,Zhang2016,Cheng2015} and WSe$_{2}$ \cite{Shao2016}, grown by chemical vapor deposition, reported nonzero spin-torque conductivities, but disagreed as to whether the dominant torque is field-like or antidamping-like.
Our research group has measured SOTs in WTe$_{2}$/permalloy samples in which semi-metallic WTe$_{2}$ layers were prepared by exfoliation, and observed an out-of-plane antidamping SOT component made possible by the low crystal symmetry of WTe$_{2}$, as well as a more-conventional in-plane anti-damping SOT and an out-of-plane field-like torque due to the Oersted field \cite{MacNeill2016,MacNeill2017}.  

For magnetic memory applications it is of particular interest to explore materials which combine high electrical conductivity, $\sigma$, and strong SOC, $\lambda _{SOC}$.
Here we report the first measurements of SOTs generated by a fully-metallic TMD, NbSe$_{2}$, with $\sigma \approx$  6 $\times$ 10$^{5}$ ($\Omega$m)$^{-1}$ in our devices and $\lambda _{SOC}$ = 76 meV \cite{Xi2015}.
For comparison, the previously-measured semiconducting TMDs MoS$_2$ and WS$_2$ have typical electrical conductivities $\sigma \approx$  10$^{-6}$ ($\Omega$m)$^{-1}$ and SOC energies $\lambda _{SOC}$ = 0 - 40 meV \cite{Liu2013,Xu2014} while for for semi-metallic WTe$_{2}$ $\sigma \approx$  10$^{4}$ ($\Omega$m)$^{-1}$ and $\lambda _{SOC}$ = 15 meV \cite{MacNeill2016,MacNeill2017,Jiang2015}.  
  
Our spin-torque ferromagnetic resonance (ST-FMR) measurements on NbSe$_{2}$/Permalloy (Py) bilayers reveal small but nonzero SOTs, corresponding to spin-torque conductivities of order or less than $10^{3} (\hbar / 2e)$($\Omega$m)$^{-1}$, about a factor of 100 weaker than the spin-torque conductivities generated by Pt or Bi$_2$Se$_3$ at room temperature \cite{MinhHai2016,Mellnik2014}.
To probe the mechanisms of these SOTs we perform systematic studies as a function of the NbSe$_{2}$ thickness, $t$, and the angle of applied magnetic field.
We measure an in-plane antidamping SOT component that is only weakly dependent of $t$, remaining sizable down to a single NbSe$_{2}$ layer -- suggesting an interfacial origin. 
We also observe a field-like out-of-plane torque that scales linearly with $t$ for sufficiently thick ($t>$ 5 nm) samples, indicating that in this regime the torque is dominated by a current-generated Oersted field.
However, for devices with smaller number of NbSe$_{2}$ layers ($t<$ 5 nm), the out-of-plane torque is  weaker than the value expected from the  field alone, and, for a single NbSe$_{2}$ layer we observe a reversal of the direction of the field-like out-of-plane torque. These deviations could be the result of either an interfacial out-of-plane field-like SOT directed opposite to the Oersted torque, or possibly to non-uniform charge current flow in the Py layer such that the current within the Py generates a nonzero net Oersted field acting on the Py.  

Interestingly, by performing systematic measurements as a function of the angle of an in-plane magnetic field we detect in some samples the presence of an additional in-plane torque that is field-like with the form $\hat{m} \times \hat{z}$, where $\hat{m}$ is the Py magnetization direction.
This torque is not allowed by symmetry considerations for the bulk NbSe$_{2}$ crystal structure \cite{MacNeill2016}.
We propose that the presence of this torque component is due to a strain-induced symmetry breaking, e.g., a unidirectional strain in the NbSe$_{2}$ layer generated during the processs of exfoliation and sample fabrication \cite{Lee2017}.

\begin{figure}[h!]
	%\centering
	\includegraphics{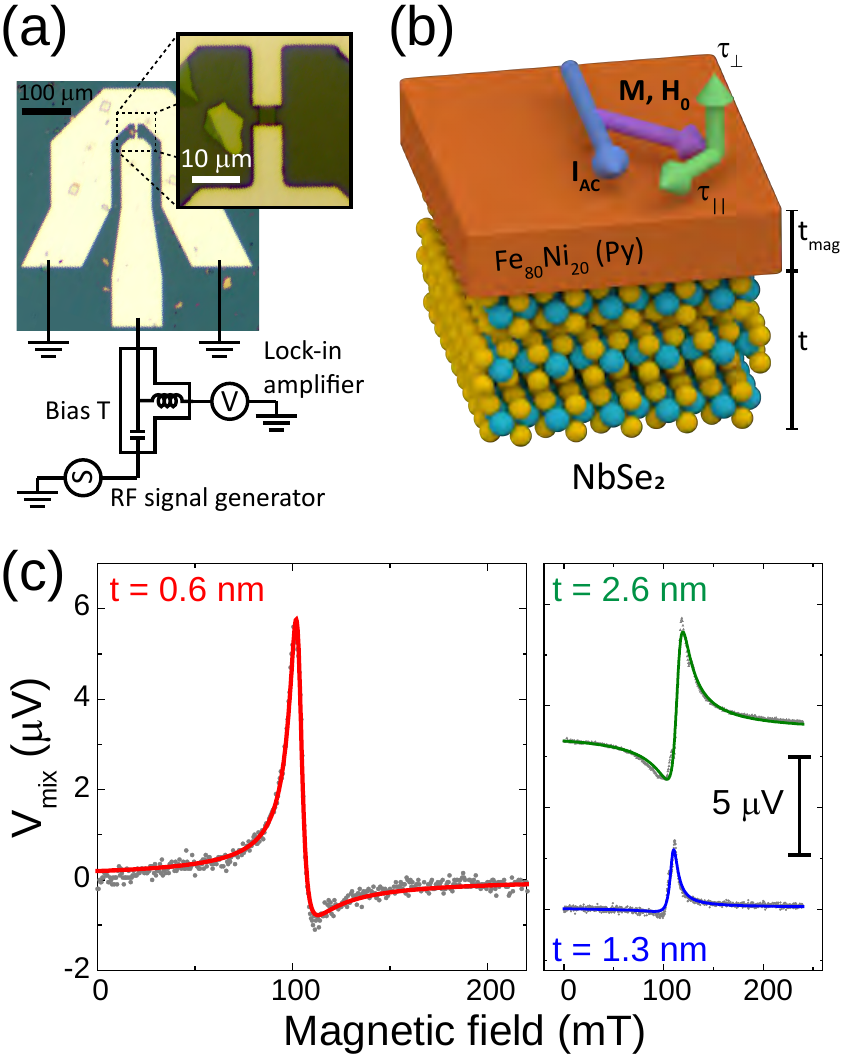}
	\caption{(a) Micrograph of a typical device with the measuring circuit schematic. (b) Schematic of the NbSe$_{2}$/Py structure. (c) ST-FMR resonances for $\phi$ = 130$^\circ$, $f$ = 9 GHz, and $P_{RF}$ = 5 dBm for different thicknesses of the NbSe$_{2}$ layer: 0.6 nm (red), 1.3 nm (blue), and 2.6 nm (green). The gray points are the measured data and the solid lines show the fits to a symmetric plus antisymmetric Lorentzian.}
	\label{fig:figure01}
\end{figure}

We prepare our samples by mechanically exfoliating a bulk synthetic NbSe$_{2}$ crystal (HQgraphene) onto an intrinsic Si wafer with a 1-$\mu$m-thick SiO$_2$ layer thermally grown on the surface.
The mechanical exfoliation is performed under vacuum (at pressures below 10$^{-6}$ Torr) in the load-lock chamber of our sputter system, and the samples are loaded into the sputtering chamber without breaking vacuum.
We then deposit 6 nm of Py by grazing angle ($\sim$ 5$^\circ$) sputtering followed by 1.2 nm of Al, which oxidizes completely upon contact with atmosphere and serves as a capping layer.
We have previously demonstrated that the grazing angle sputter deposition causes little to no damage to our TMD crystals \cite{MacNeill2016}.
The NbSe$_{2}$ flakes are identified by optical contrast, and their thicknesses and morphology are determined by atomic force microscopy (AFM).
In order to avoid artifacts in our measurements due to roughness of the ferromagnetic layer, we selected only flat flakes with RMS surface roughness below 0.4 nm, measured by AFM in an area of 1x1 $\mu$m$^{2}$, and with no steps in the TMD crystals over the device area.
We then pattern the NbSe$_{2}$/Py bilayers into a bar shape with a well-defined length and width by using electron beam lithography followed by Ar$^{+}$ ion milling.
As a final step, we define Ti/Pt (5/75 nm) contacts in the shape of a coplanar waveguide using electron beam lithography followed by metal sputtering deposition.
An optical micrograph of a finished device is shown in Figure \ref{fig:figure01}a.

To measure the SOTs, we use the ST-FMR technique \cite{Liu2011,Mellnik2014,MacNeill2016} in which an alternating microwave-frequency current ($I_{RF}$) (with frequencies f  = 7 - 12 GHz) is applied within the sample plane.
Current-induced torques cause the magnetization $M$ of the ferromagnet to precess.
By applying an external magnetic field $H$ at an angle $\phi$ with respect to $I_{RF}$, we set the direction of $\vec{M}$ and the characteristic ferromagnetic resonance frequency of the ferromagnet (Figure \ref{fig:figure01}b).
The precession of the magnetization creates a time-dependent change of the device resistance due to the anisotropic magnetoresistance (AMR) of the ferromagnetic layer.
This change in resistance mixes with $I_{RF}$ generating a DC voltage across the NbSe$_{2}$/Py bar (V$_{mix}$).
The signal-to-noise ratio is maximized by modulating the amplitude of $I_{RF}$ at low frequencies and detecting V$_{mix}$ using a lock-in amplifier.
The circuit geometry is shown in Figure \ref{fig:figure01}a.
All measurements are performed at room temperature unless indicated.

When the ferromagnetic resonance frequency matches $f$, V$_{mix}$ shows a resonance peak with a lineshape that can be described as: $V_{mix}(H) = V_{S}(H) + V_{A}(H)$, where $V_{S}$ is a symmetric Lorentzian with amplitude proportional to the in-plane component of the torque ($\tau_{\parallel}$), and $V_{A}$ an antisymmetric Lorentzian with amplitude proportional to the out-of-plane component of the torque ($\tau_{\perp}$).
This allows the separation of the two torque components by fitting a measurement of V$_{mix}$ as a function of $H$.
The two torques components are related to the amplitudes of the Lorentzians by \cite{Liu2011, Mellnik2014}:

\begin{equation}
  V_{S} = - \frac{I_{RF}}{2} \left( \frac{dR}{d \phi} \right) \frac{1}{\alpha \gamma \mu_{0} \left( 2 H_{0} + M_{eff} \right)} \tau_{\parallel},
	\label{eq:VS}
\end{equation}

\begin{equation}
  V_{A} = - \frac{I_{RF}}{2} \left( \frac{dR}{d \phi} \right) \frac{\sqrt{1+ M_{eff}/H_{0}}}{\alpha \gamma \mu_{0} \left( 2 H_{0} + M_{eff} \right)} \tau_{\perp},
	\label{eq:VA}
\end{equation}

\noindent where $R$ is the device resistance, $\phi$ is the angle between $H$ and $I_{RF}$, $M_{eff}$ is the effective magnetization of the Py layer, composed by the saturation magnetization and anisotropy terms, $\alpha$ is the Gilbert damping of the Py, $\gamma$ is the gyromagnetic ratio, $\mu_{0}$ is the vacuum permittivity, and $H_{0}$ is the resonance field.
The term $dR/d \phi$ is due to the AMR in the Py layer.
For our devices we have $\mu_{0} M_{eff} \approx$  0.8 T and $\alpha \approx$ 0.01 as obtained by the ST-FMR resonance frequency and linewidth, respectively, and $R ( \phi )$ is measured directly by measuring the device’s resistance as a function of $\phi$.
The current $I_{RF}$ is calibrated by using a network analyzer to measure transmitted and reflected microwave powers ($S_{11}$ and $S_{21}$).

Resonance curves for one, two, and four NbSe$_{2}$ monolayers devices ($t$ = 0.6, 1.3, and 2.6 nm, respectively) with $f$ = 9 GHz, applied RF power $P_{RF}$ = 5 dBm and $\phi$ = 130$^\circ$ are shown in Fig.~\ref{fig:figure01}c, where the gray points represent experimental data and the fits are shown by the solid lines.
Two important features are illustrated by these curves: the ratio between the amplitude of the symmetric and antisymmetric Lorentzians decreases with the increase of $t$, and the sign of the antisymmetric component flips sign between the mono- and bi- layer devices.
For both the one and two layer-thick devices, the lineshape is dominated by the symmetric component of the Lorentzian, meaning that the in-plane SOT is dominant over the out-of-plane component.

Our observation of both symmetric and antisymmetric components in the ST-FMR resonance is qualitatively similar to the results of Ref.~6 on MoS$_{2}$/Py devices. The presence of both field-like and damping-like interfacial torques is consistent with general considerations of interfacial spin-orbit torques \cite{Stiles2016F, Stiles2016P, Stiles2017}.
However, a more recent measurement on MoS$_2$/CoFeB and WSe$_2$/CoFeB structures using a second harmonic Hall technique was unable to measure any in-plane SOT and attributed the large symmetric Lorentzian measured in Ref.~6 to spin-pumping combined with an inverse Rashba-Edelstein effect, rather than a spin-orbit torque \cite{Shao2016}.  We can tell that the symmetric ST-FMR resonance signal we observe is not due primarily to a spin-pumping effect because this would require an unphysically-large spin-to-charge conversion factor (see Supplementary Information).

The symmetries and mechanisms of the SOTs can be analyzed in more detail by performing ST-FMR measurements as a function of the magnetic-field angle as $\tau_{\parallel}$ and $\tau_{\perp}$ both depend upon $\phi$.
The contributions to the expected angular dependence can be understood as follows.
Part of the angular dependence arises from the AMR in the bilayer, which contributes the dependence $dR / d \phi \propto \sin(2 \phi )$ (see Eqs.\ \ref{eq:VS} and \ref{eq:VA}).
Many current-induced torques have a $\cos( \phi )$ dependence (e.g. in-plane antidamping torques due to standard spin Hall or Rashba-Edelstein Effects, and the field-like out-of-plane torque due to the Oersted field), leading to an overall angular dependence $V_{mix} \propto \cos( \phi ) \sin(2 \phi )$.
However, additional torque components can arise in systems with lower-symmetry \cite{Garello2013,Yu2014}, such as WTe$_{2}$ \cite{MacNeill2016,MacNeill2017} and some semiconductor alloys \cite{Chernyshov2009,Endo2010,Fang2011,Skinner2015,Ciccarelli2016}.

\begin{figure*}[ht!]
		\includegraphics[scale=1.0]{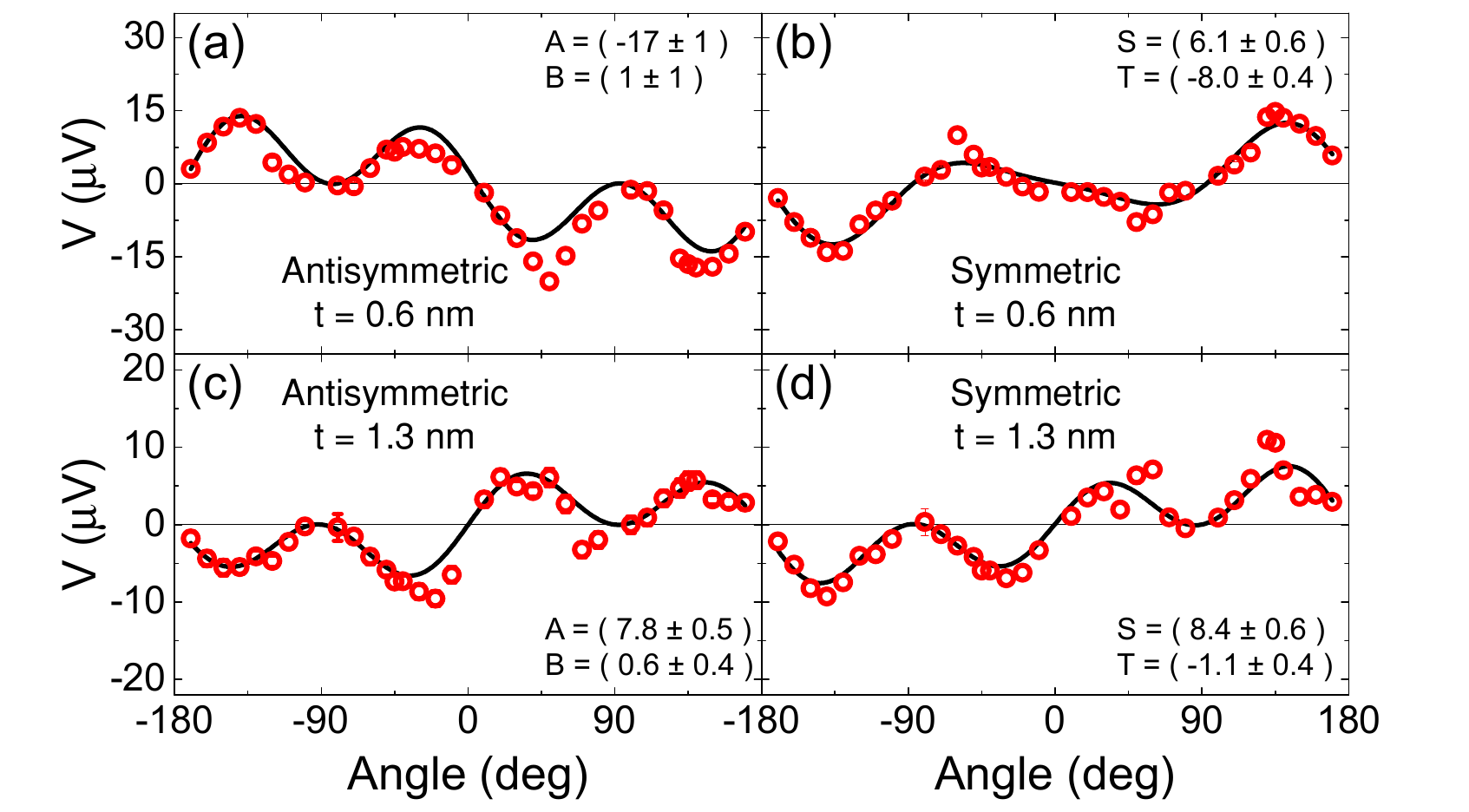}
	\caption{Antisymmetric and symmetric components of the ST-FMR resonance fits for $V_{mix}$ as a function of the magnetic field angle for devices with (a, b) one and (c, d) two NbSe$_{2}$ monolayers for $f$ = 9 GHz and $P_{RF}$ = 5 dBm. The data are shown by the red circles and the fits using equations \ref{eq:VSfit} and \ref{eq:VAfit} are shown by the black lines.}
	\label{fig:figure02}
\end{figure*}

The angular dependence we measure for the antisymmetric and symmetric components of the ST-FMR resonances of NbSe$_{2}$/Py samples are shown in Fig. \ref{fig:figure02} for devices with monolayer (a,b) and bilayer (c,d) NbSe$_{2}$. 
The angular dependence of the antisymmetric components for both samples is consistent with a simple $\cos( \phi ) \sin(2 \phi )$ form, illustrating that the out-of-plane torque has the usual $\cos( \phi )$ dependence expected for a field-like out-of-plane torque.
However, the symmetric ST-FMR components deviate from this form.
We have performed more general fits (black lines in Fig. \ref{fig:figure02}) for both components to the forms:

\begin{equation}
  V_{S} = S \cos \left( \phi \right) \sin \left( 2 \phi \right) + T \sin \left( 2 \phi \right),
	\label{eq:VSfit}
\end{equation}

\begin{equation}
  V_{A} = A \cos \left( \phi \right) \sin \left( 2 \phi \right) + B \sin \left( 2 \phi \right),
	\label{eq:VAfit}
\end{equation}

\noindent corresponding to the inclusion of additional angle-independent torques $\tau_{T} \propto T$ and $\tau_{B} \propto B$ such that the in-plane torque is $\tau_{\parallel} = \tau_{S} \cos( \phi) + \tau_{T}$ and the out-of-plane torque is $\tau_{\perp} = \tau_{A} \cos( \phi) + \tau_{B}$ (where $\tau_{S}$, $\tau_{T}$, $\tau_{A}$, and $\tau_{B}$ are independent of $\phi$).
The vector forms of these additional torque components correspond to $\vec{\tau_{T}} \propto \hat{m} \times \hat{z}$ and $\vec{\tau_{B}} \propto \hat{m} \times \left( \hat{m} \times \hat{z} \right)$.
We find that this generalization greatly improves the fits for the symmetric ST-FMR components, with nonzero values for both $S$ and $T$, and with the results for the monolayer sample indicating $\left| \tau_{T} \right| > \tau_{S}$.
For the bilayer sample the contribution from $\tau_{T}$ is less prominent but still clearly nonzero, while for both samples the fits to the antisymmetric component gives $\tau_{B}$ = 0 within the experimental resolution.  

This result is curious in several ways.
First, for the usual 2H-NbSe$_{2}$ structure (space group P6$_{3}$/mmc) \cite{Jaque2014}, the NbSe$_{2}$/Py interface reduces to the space group P3m1 containing the identity, two 3-fold rotations, and three mirror planes.
This set of symmetries forbids the presence of both of the torque terms $\vec{\tau_{T}} \propto \hat{m} \times \hat{z}$ and $\vec{\tau_{B}} \propto \hat{m} \times \left( \hat{m} \times \hat{z} \right)$.
However, any uniaxial strain will break the three-fold rotational symmetry and reduce the mirror symmetries to a single mirror plane or lower, depending on the alignment of the strain axis to the crystal axes. If there is a uniaxial strain, the torque terms $\propto \hat{m} \times \hat{z}$ or $\propto \hat{m} \times \left( \hat{m} \times \hat{z} \right)$ become symmetry-allowed, and in the case of a remaining mirror plane, the applied electrical current must have a component perpendicular to this plane.
This situation is analogous to the strain-induced valley magnetoelectic effect observed in MoS$_{2}$ monolayers \cite{Lee2017}.
 
%%%
%Our estimates of $\tau_{T}$'s dependence on the current flow direction with respect to the NbSe$_2$ crystal axes are suggestive of this mechanism (Supplemental Information).
%%%

We note, though, that both $\vec{\tau_{T}}$ and $\vec{\tau_{B}}$ are subject to the same symmetry constraints, so it is curious that $\vec{\tau_{B}}$ remains zero even when broken symmetries allow $\vec{\tau_{T}} \neq 0$.
Furthermore, the result we find in the (presumably strained) NbSe$_{2}$/Py samples ($\vec{\tau_{T}} \neq 0$, $\vec{\tau_{B}} = 0$) is opposite to the results in WTe$_{2}$/Py samples ($\vec{\tau_{T}} = 0$, $\vec{\tau_{B}} \neq 0$) \cite{MacNeill2016,MacNeill2017} where a similar low-symmetry state is intrinsic to the WTe$_{2}$ crystal structure.
This suggests that the existence of the torque components $\vec{\tau_{B}}$ and $\vec{\tau_{T}}$ depend on more than the nature of the global broken symmetries, but also on microscopic factors like the nature of the atomic orbitals that contribute to charge and spin transport.
%Additionally, while for WTe$_{2}$ we observed a clear magnetic anisotropy term correlated with the crystalline axes \cite{MacNeill2016}, we did not observe such a term in any of our NbSe$_{2}$ devices (see Supplementary Information). {\bf[This does not actually seem to be mentioned in the SI.  If you decide to keep the discussion here, you should mention what is the minimum value of anisotropy to which you are sensitive.  You could also just delete this sentence about anisotropy]}
%\hl{[This is perhaps not surprising as there is no in-plane symmetry breaking direction per the Haney mechanism, and is only (presumably) weakly broken in the strained samples.]}

\begin{figure}[h]
		\includegraphics[width=0.42\textwidth]{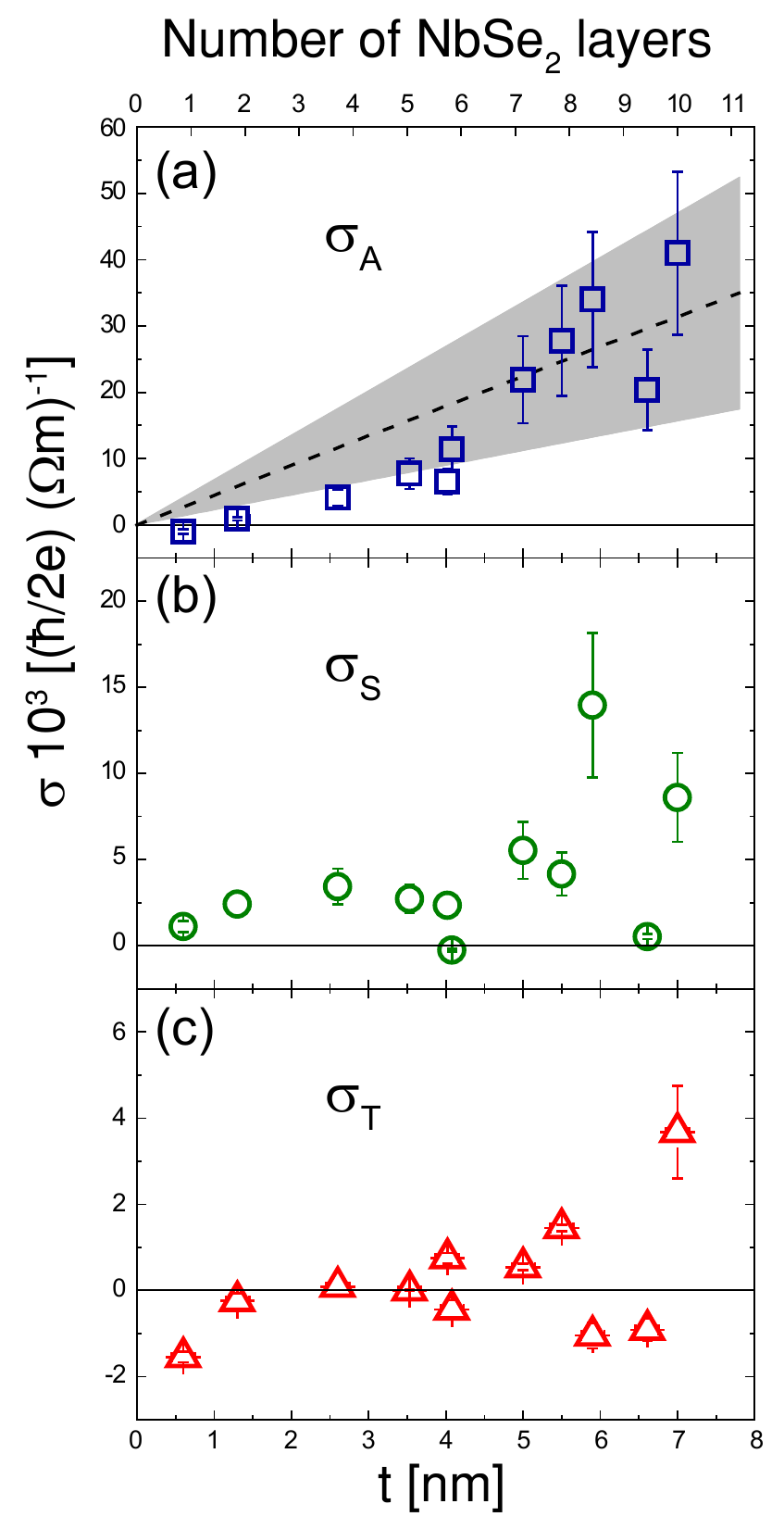}
	\caption{Spin-torque conductivities (a) $\sigma_{A}$, (b) $\sigma_{S}$, and (c) $\sigma_{T}$ as a function of the NbSe$_{2}$ thickness obtained from the angular fits for $f$ = 9 GHz and $P_{RF}$ = 5 dBm. The dashed line in (a) shows the estimated contribution from the Oersted field ($\sigma_{Oe}$) with the gray area representing its standard deviation.}
	\label{fig:figure03}
\end{figure}

We investigated the extent to which the different torque components $\tau_{S}$, $\tau_{T}$, $\tau_{A}$, and $\tau_{B}$ depend on the NbSe$_{2}$ thickness, $t$, by performing ST-FMR measurements as a function of applied magnetic field angle for a collection of different devices with different values of $t$, while keeping the Py thickness fixed ($t_{mag}$ = 6 nm).
These strengths of each torque component are linear in the current and voltage applied to the sample, and because the electric field across the device can be more accurately determined than the separate current densities through each of the individual layers in our devices (NbSe$_{2}$ and Py), we express the torque strengths as torque conductivities, $\sigma_{j} \equiv d \tau_{j} / dE$ in units of ($\hbar / 2e$)($\Omega m$)$^{-1}$, where $j$ = $A$, $B$, $S$, or $T$ corresponds to the different torque components, $E$ is the electric field, $\hbar$ is the reduced Plank's constant and $e$ the electron charge.
We plot the thickness dependence of $\sigma_{A}$, $\sigma_{S}$, and $\sigma_{T}$ in Fig. \ref{fig:figure03}.
The component $\sigma_{B}$ is zero within experimental error for all of the samples measured.

For the out-of-plane field-like torque conductivity $\sigma_{A}$ we observe a clear increase with increasing NbSe$_{2}$ thickness (Fig. \ref{fig:figure03}a).
For our thicker devices the magnitude of $\sigma_{A}$ agrees with our estimation of the Oersted-field contribution ($\sigma_{Oe}$) due to the current flowing in the NbSe$_{2}$ layer.
However, for the thinner ($t<$ 5 nm) devices, $\sigma_{A}$ is significantly lower than our estimate for the Oersted contribution $\sigma_{Oe}$, and then the sign of $\sigma_{A}$ is reversed for the monolayer device (see Fig.\ \ref{fig:figure01}c).
This behavior at small NbSe$_{2}$ thicknesses suggests the presence of an interfacial field-like SOT that opposes the Oersted contribution.
However, the size of the reversed SOT is sufficiently small that it is difficult to rule out possible alternative mechanisms such as a spatially non-uniform current density through the thickness of the Py layer. (Nonzero antisymmetric ST-FMR resonances can be observed even in single-layer Py samples, and have been ascribed to this mechanism \cite{Liu2011, Emori2016}.)

The in-plane damping-like torque component $\sigma_{S}$ (Fig. \ref{fig:figure03}b) has at most a weak dependence on $t$, and possesses a non-zero value all the way down to a single NbSe$_2$ layer.
The small apparent increase of $\sigma_{S}$ with increasing $t$ could arise from a bulk contribution, such as the spin Hall effect.
However, the nonzero value of this term down to a single NbSe$_2$ layer suggests a sizable interfacial SOT.
The value of $\sigma_{S}$ for the thinnest samples ($\sigma_{S} \approx$ 3 $\times 10^{3} \left( \hbar / 2e \right) \left( \Omega m \right) ^{-1}$) has a magnitude similar to reports for other TMDs, such as MoS$_{2}$ \cite{Zhang2016} and WTe$_{2}$ \cite{MacNeill2016}, but it is significantly below the values for Pt/ferromagnet bilayers \cite{MinhHai2016} and topological insulators at room temperature \cite{Mellnik2014}  ($\sigma_{S} \approx$ $10^{5} \left( \hbar / 2e \right) \left( \Omega m \right) ^{-1}$).

For the in-plane field-like torque $\sigma_{T}$ that is forbidden by symmetry for unstrained NbSe$_2$ (Fig. \ref{fig:figure03}c), we do not observe any systematic trend in the torque conductivity as a function of $t$.
While $\sigma_{T} \approx 0$ for a few of our devices, both the sign and magnitude of this $\hat{m} \times \hat{z}$ torque term seem uncorrelated with the thickness of the NbSe$_{2}$ layer.
The lack of correlation between $\sigma_{T}$ and $t$ is in agreement with our assumption that this term arises due to strain in our samples since we do not control this parameter.
Strain-controlled experiments \cite{Lee2017} could be performed in order to confirm this assumption and better constrain the microscopic origin of this extra torque term.

We also performed temperature dependence measurements for a sample showing all three torque components: $\sigma_{A}$, $\sigma_{S}$, and $\sigma_{T}$ (see Supplementary Information), with $\sigma_{A} \approx \sigma_{Oe}$.
We observe only a weak temperature dependence for the torque ratio $\sigma_{S} / \sigma_{A}$ and a slightly stronger temperature dependence for $\sigma_{T} / \sigma_{A}$.
The weak temperature dependence of the interfacial SOTs in TMD/ferromagnet bilayers is in agreement with previous studies on semiconducting TMDs \cite{Shao2016}.

%%%%%%%%%%%%%%%%%%%%%%%%%%%%%%%%%%%%%%%%%%%%%%%%%%%%%%%%%%%%%%%%%%%%%%%%%%%%%%%%%%%%%%%
%\section{Conclusions}
In summary, we report current induced SOTs in NbSe$_{2}$/Py bilayers.
The in-plane antidamping-like term has only a very weak dependence with $t$, with values for the spin torque conductivity comparable to other TMDs.
For thin NbSe$_{2}$ layers, the out-of-plane SOT component for thin NbSe$_{2}$ layers is significantly smaller than the estimate Oersted-field contribution, with a sign reversal for a monolayer of NbSe$_{2}$.
In additional to these expected torque components, we also observe the presence of a SOT with the form $\hat{m} \times \hat{z}$ which is forbidden by the bulk symmetry of the NbSe$_{2}$ crystal, but can arise in the presence of strain.

%%%%%%%%%%%%%%%%%%%%%%%%%%%%%%%%%%%%%%%%%%%%%%%%%%%%%%%%%%%%%%%%%%%%%%%%%%%%%%%%%%%%%%%
%\section{Acknowledgments}
We thank R.~A.~Buhrman for valuable discussions and comments on the manuscript.
This work was supported by the Kavli Institute at Cornell for Nanoscale Science, the Netherlands Organization for Scientific Research (NWO Rubicon 680-50-1311), the National Science Foundation (DMR-1406333, DMR-1708499), and the Army Research Office (W911NF-15-1-0447).
G.M.S. acknowledges support by a National Science Foundation Graduate Research Fellowship under Grant No. DGE-1144153.
This work made use of the NSF-supported Cornell Nanoscale Facility (ECCS-1542081) and the Cornell Center for Materials Research Shared Facilities, which are supported through the NSF MRSEC Program (DMR-1719875).

%%%%%%%%%%%%%%%%%%%%%%%%%%%%%%%%%%%%%%%%%%%%%%%%%%%%%%%%%%%%%%%%%%%%%%%%%%%%%%%%
\bibliography{nbse2-ref}

\end{document}